\documentclass[12pt]{article}
\usepackage{graphicx}
\usepackage{pdfpages}

\usepackage[margin=1in]{geometry}
\usepackage{amsmath,amssymb,amsfonts}
\usepackage{mathtools}
\usepackage{physics}
\usepackage{hyperref}
\usepackage{cleveref}
\usepackage{graphicx}
\usepackage{booktabs}
\usepackage{microtype}
\usepackage{enumitem}
\usepackage{float}

\usepackage{amssymb}
\usepackage{amsmath}
\usepackage{cancel}
\usepackage{color}
\usepackage{float}
\usepackage{enumitem}
\usepackage{relsize}
\usepackage[normalem]{ulem}
\usepackage[dvipsnames]{xcolor}
\usepackage[normalem]{ulem}
\usepackage[numbers,sort&compress]{natbib}
\usepackage{hyperref}

\newcommand{\be}{\begin{equation}}
\newcommand{\ee}{\end{equation}}

\begin{document}
\numberwithin{equation}{section}

\begin{center}

{\Large \bf{ Wavefunction  Collapse 
%\\\vspace{3mm} 
in String Theory
%\\\vspace{3mm}
%from Instant Folded Strings
}}

\vspace{12mm}

%\break \break\break
\textit{Nissan Itzhaki}
\break \break\break
  School of Physics and Astronomy, Tel Aviv University, Ramat Aviv, 69978, Israel
 
\end{center}

%\date{\today}
\vspace{5mm}

\begin{abstract}

One of the most intriguing proposals for wavefunction collapse is the Diósi–Penrose model, in which collapse is driven by stochastic fluctuations of the Newtonian potential. We argue that a closely related effective structure can emerge in string theory if, as recently suggested, the present cosmic acceleration is sourced by instant folded strings and their decay products. A key difference, however, is that in this stringy setting the noise is naturally colored in time rather than white. As a result, the scenario is significantly less constrained by existing experiments than the standard Diósi–Penrose model.

\end{abstract}

\newpage

\baselineskip=18pt

\section{Introduction}

The measurement problem in quantum mechanics has puzzled physicists for nearly a century. 
Since Zeh's 1970 paper \cite{Zeh1970}, it has become increasingly clear that environmental entanglement and decoherence play a central role in any realistic account of measurement: by suppressing interference terms, they render the reduced density matrix approximately diagonal in the pointer basis.\footnote{Earlier precursors of the broader idea of interference suppression, rather than decoherence in the sense associated with the measurement problem, can be traced all the way to the late 1920s. Mott \cite{Mott1929}, for example,
showed that a classical-looking particle trajectory can emerge from a purely quantum wave through the entangled correlations between the particle and the detector medium.}
Whether decoherence by itself suffices to solve the measurement problem, however, remains a matter of debate \cite{Zurek2003,Schlosshauer2005}. Very broadly speaking, two main lines of thought have emerged.

In the first, usually associated with the Everettian or many worlds interpretation \cite{Everett1957}, the wavefunction never collapses, but always evolves unitarily according to the Schr\"odinger equation. On this view, what is conventionally called a measurement outcome corresponds to the observer and apparatus becoming entangled with the measured system, so that the global state evolves into a superposition of effectively non-interacting branches, each associated with a different result. Decoherence rapidly suppresses interference between these branches, thereby explaining why observers within any given branch perceive a single definite effectively classical outcome.

The second broad approach is the objective collapse, or dynamical reduction, program, in which wavefunction collapse is promoted to a genuine physical process through stochastic and nonlinear modifications of Schr\"odinger evolution. In such models, macroscopic superpositions are dynamically suppressed, and measurement-like amplification processes drive the state toward a single definite outcome, while standard quantum predictions are recovered to an excellent approximation at the microscopic level. Different proposals differ mainly in their choice of collapse operators, noise field, and amplification mechanism, which together determine the preferred basis and collapse rate. In particular, modern objective collapse models are typically observer independent: the reduction process is taken to occur spontaneously or through a physical mechanism, such as mass density amplification or gravity related effects, rather than being triggered by an external observer or by an ill-defined notion of ``measurement'' \cite{GRW1986,Pearle1989,GPR1990,Diosi1989,Penrose1996,BassiGhirardi2003,BassiEtAl2013,BassiDoratoUlbricht2023}.

A key virtue of objective collapse models is that they are experimentally testable. Generically, they predict small departures from standard quantum mechanics, often captured phenomenologically by an additional universal noise that suppresses spatial coherence and can also induce momentum diffusion, excess heating, and, in some realizations, spontaneous radiation \cite{BassiGhirardi2003,BassiEtAl2013,BassiDoratoUlbricht2023}. Such effects can be probed both interferometrically, most directly in high mass matter wave interferometry, and non interferometrically, through optomechanical platforms and precision measurements searching for anomalous force noise or heating \cite{Fein:2019dgf,CarlessoPaternostro2020,VinanteEtAl2016,CarlessoEtAl2022}. Dedicated X-ray/$\gamma$-ray searches provide a further handle on collapse mechanisms that predict spontaneous emission \cite{Fu1997,ArnquistEtAl2022,XENON:2025plg}. As experiments achieve coherent control over increasingly massive systems, larger delocalizations, and longer coherence times, the continued absence of deviations from standard quantum mechanics translates into progressively stronger bounds on the collapse parameters, whereas a positive signal would point to a genuine breakdown of purely unitary quantum dynamics \cite{BassiEtAl2013,Fein:2019dgf,CarlessoEtAl2022}.

\begin{figure}[t!]
\centering
\includegraphics[width=16cm]{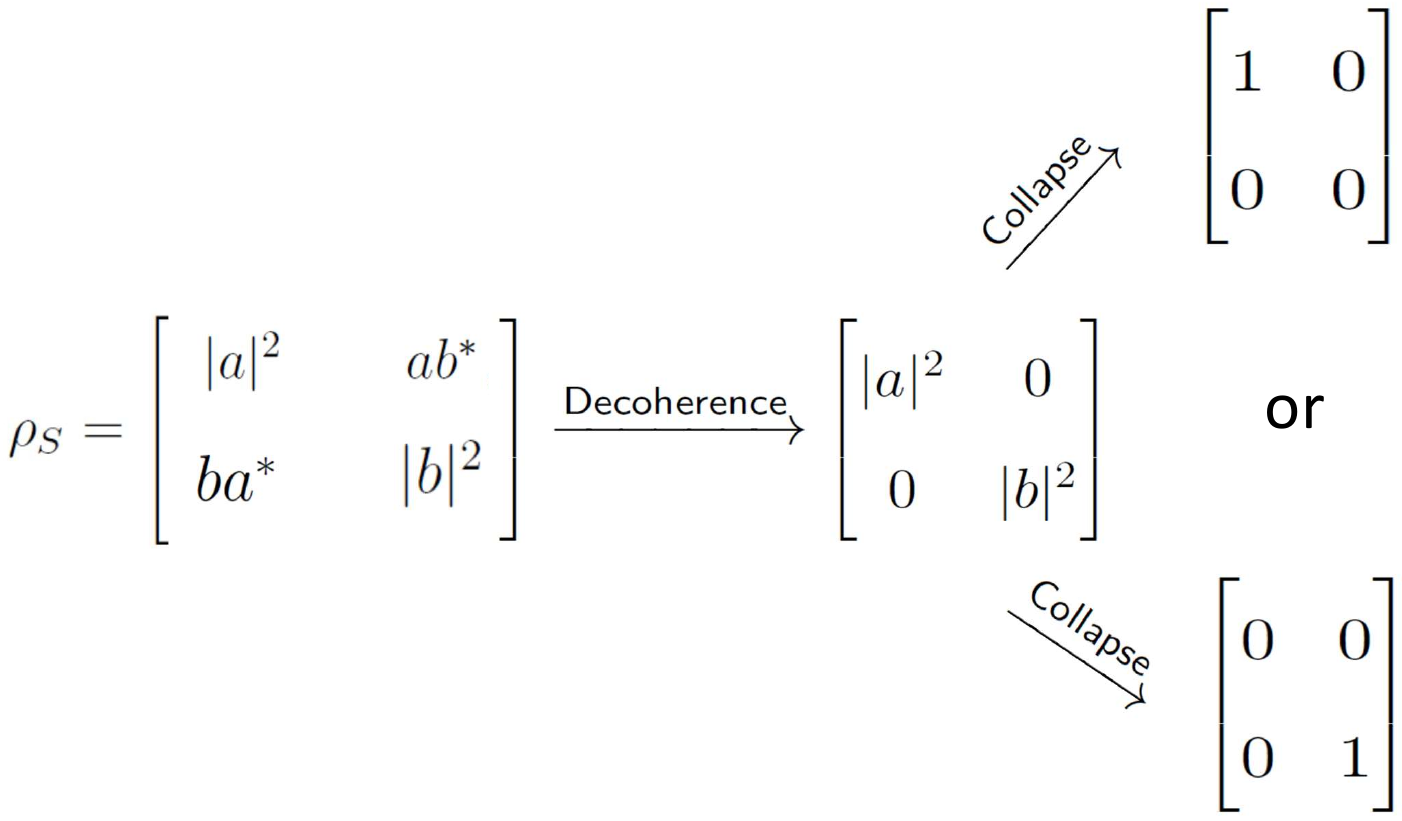}
\vspace{-2cm}
\caption{Decoherence approximately diagonalizes the density matrix. In the many-worlds approach, this is the end of the story: the different branches persist without further collapse. In objective collapse models, by contrast, an additional non-unitary evolution selects one definite outcome.}
\label{fig:decoherence_branches}
\end{figure}

Among the various objective collapse proposals, one of the best studied gravity related models is the Di\'osi--Penrose (DP) model. In Di\'osi's formulation, collapse is implemented through a stochastic modification of Schr\"odinger dynamics, while Penrose argued, more heuristically, that spatial superpositions of different mass distributions should possess a finite lifetime set by their gravitational self energy. The appeal of such proposals is twofold. On the one hand, they provide a concrete phenomenological arena in which the measurement problem may be linked to gravity. On the other hand, once formulated as dynamical models, they predict observable departures from standard quantum mechanics and are therefore subject to experimental scrutiny.

String theory does not appear to offer support for objective-collapse modifications of quantum mechanics. In the settings that are theoretically best understood, it instead seems to preserve the standard quantum mechanical framework: in asymptotically flat space, perturbative string theory is organized around conventional $S$-matrix amplitudes, while in asymptotically $AdS$ space holography provides a non-perturbative definition in terms of a quantum field theory living at the boundary \cite{Maldacena1998,GKP1998,Witten1998}.

We, however, do not live in either $AdS$ or Minkowski space, and the situation is considerably less settled in de~Sitter-like backgrounds, where both the construction of controlled string vacua and the identification of precise observables remain substantially more subtle
\cite{Witten2001,DanielssonVanRiet2018,DineEtAl2021,CicoliEtAl2024}.
More generally, formulating string theory in time-dependent settings has proven to be a challenging task, one that is likely to require nontrivial conceptual developments.
It is therefore conceivable that, in cosmological backgrounds, string theory could provide a mechanism for objective collapse.

The aim of this note is to argue that, at least in a specific cosmological setting, this is indeed the case. We will claim that, in the framework considered in \cite{Itzhaki:2024pok}, in which the present cosmic acceleration is attributed to Instant Folded Strings (IFSs) \cite{Itzhaki:2018glf} and their decay products, string theory gives rise to a rather precise objective collapse mechanism whose structure closely resembles the DP-model. The basic claim is that the same stringy sector that causes the universe to accelerate also sources long wavelength gravitational fluctuations with a DP-like structure.
This may be viewed as a microscopic, UV-complete realization of the DP model within string theory.

The paper is organized as follows. Section~2 reviews the DP model, and Section~3 discusses the corresponding experimental bounds. In Section~4 we introduce a toy model that reproduces DP-like dynamics, and in Section~5 we show that the stringy dark energy setup of \cite{Itzhaki:2024pok} provides a UV-complete description of this toy model. Section~6 is devoted to experimental constraints, while Section~7 contains the discussion.

\section{The Diósi--Penrose model}
\label{sec:DP_model}
The basic idea is to use gravity to construct a model of wavefunction collapse. This is a rather natural proposal. Gravity is central to the dynamics of macroscopic objects, yet it is typically negligible for microscopic ones, and so it is tempting to ask whether gravity itself could trigger collapse.

Penrose's starting point is the tension between (i) the quantum superposition principle and (ii) the equivalence principle of GR.
A massive body localized in two well separated positions is, in semiclassical gravity, associated with two different approximately stationary space--time geometries. A superposition of the corresponding mass configurations, therefore, suggests a superposition of two space--time geometries. He argues that such a superposition is fundamentally ``ill-defined'' because, in general relativity, time translations are generated by the space--time geometry itself. In a superposition of geometries, there is therefore no unique time translation operator, and the notion of a stationary superposed state becomes ambiguous.

Penrose further argued \cite{Penrose1996,Penrose1998} that this ambiguity signals an instability that leads to wavefunction collapse. More specifically, the ill-definedness is assumed to behave like an effective energy uncertainty, $\Delta E_G$, associated with the mismatch between the two gravitational fields. Time energy uncertainty then implies
\be
\label{eq:Penrose_tau}
\tau_{\rm OR} \;\sim\; \frac{\hbar}{\Delta E_G},
\ee
is a natural timescale for objective reduction. For quasi-static mass distributions $m_1(\mathbf{x})$ and $m_2(\mathbf{x})$ corresponding to the two branches, a commonly used expression for $\Delta E_G$ is the Newtonian gravitational self-energy of the difference between the two mass densities,
\begin{equation}
\label{eq:Penrose_DeltaEG}
\Delta E_G \;\equiv\; \frac{G}{2}\int d^3x\, d^3y\;
\frac{\big(m_1(\mathbf{x})-m_2(\mathbf{x})\big)\,\big(m_1(\mathbf{y})-m_2(\mathbf{y})\big)}
{\lvert \mathbf{x}-\mathbf{y}\rvert}\,,
\end{equation}
up to order one convention choices.\footnote{Different papers place the overall sign and some factors of $1/2$ slightly differently; these conventions do not affect the central scaling $\tau\sim\hbar/\Delta E_G$.}

As far as we know, Penrose did not commit to a specific dynamical model. Diósi, however, did. In fact, a few years before Penrose's proposal, Diósi introduced a concrete dynamical model \cite{Diosi1987}, now commonly referred to as the DP-model. Its first ingredient is the standard non relativistic gravitational coupling,
\begin{equation}
\label{eq:H_int}
H_{\rm int}(t)=\int d^3x\,\hat{m}(\mathbf{x})\,\Phi(\mathbf{x},t),
\end{equation}
where $\hat m$ is the mass density operator and $\Phi$ is the gravitational potential. Diósi proposed that $\Phi$ contains a classical stochastic component that is white in time and falls off as $1/r$:
\begin{equation}
\label{eq:DP_Phi_corr_real}
\mathbb{E}\!\left[\Phi(\mathbf{x},t)\,\Phi(\mathbf{y},t')\right]
\;=\; \hbar G\;\frac{1}{\lvert \mathbf{x}-\mathbf{y}\rvert}\;\delta(t-t'),
\end{equation}
or, equivalently, in Fourier space (with $p=(\omega,\mathbf{k})$),
\begin{equation}
\label{eq:DP_Phi_corr_fourier}
\mathbb{E}\!\left[\Phi(p)\,\Phi(p')\right]
\;=\;
(2\pi)^4\delta^{(4)}(p+p')\;
\hbar G\,\frac{4\pi}{k^2}\,.
\end{equation}
Namely, the standard DP spatial spectrum is $1/k^2$ together with a flat frequency spectrum.
In colored-in-time variants, one replaces $\delta(t-t')$ by a short-range kernel $g(t-t')$, which corresponds to multiplying the right hand side of \cref{eq:DP_Phi_corr_fourier} by a frequency dependent factor $\tilde g(\omega)$. As we shall see, this generalization is relevant to the string theoretic setting, which is not surprising given that string theory is relativistic.

The stochastic component of $\Phi$ induces a non-unitary evolution in which spatial superpositions of distinct mass distributions are progressively suppressed. The simplest way to describe this is through the evolution equation for the density matrix, which now takes the form
\begin{equation}
\label{eq:DP_master}
\frac{d\hat{\rho}}{dt}
=
-\frac{i}{\hbar}[\hat{H},\hat{\rho}]
-\frac{G}{2\hbar}\int d^3x\,d^3y\;
\frac{[\hat{m}(\mathbf{x}),[\hat{m}(\mathbf{y}),\hat{\rho}]]}{\lvert \mathbf{x}-\mathbf{y}\rvert}\,,
\end{equation}
where the additional double commutator term suppresses off diagonal elements in a basis of macroscopically distinct mass density configurations. In particular, for a superposition of two quasi classical mass distributions (branches $1,2$), the DP decoherence rate is controlled by the same gravitational self energy functional $\Delta E_G$ that appears in Penrose's argument (\ref{eq:Penrose_tau}), leading to a characteristic decay time $\tau\sim\hbar/\Delta E_G$ up to order one factors \cite{Diosi1987,Penrose1996,Donadi2021}.

The Newtonian kernel together with the operator $\hat{m}(\mathbf{x})$ makes the model UV sensitive. If $\hat{m}(\mathbf{x})$ is taken to be a sum of pointlike nucleon densities, the model predicts enormous momentum diffusion (``spontaneous heating'') and, for charged constituents, spontaneous radiation emission \cite{Diosi2014Bulk,Donadi2021}. For this reason, essentially all phenomenological applications introduce a finite smearing length (often denoted $R_0$ or $\sigma_{\rm DP}$) by replacing $\hat{m}(\mathbf{x})$ with a coarse-grained mass density,
\begin{equation}
\label{eq:smearing}
\hat{m}_{R_0}(\mathbf{x})
=
\int d^3z\;
g_{R_0}(\mathbf{x}-\mathbf{z})\,\hat{m}(\mathbf{z}),\qquad
g_{R_0}(\mathbf{r})
=
\frac{1}{(2\pi R_0^2)^{3/2}}\,e^{-\mathbf{r}^2/(2R_0^2)}\,,
\end{equation}
or by using a similar UV regulator. The parameter free ``natural'' choice, $R_0\sim 10^{-15}\,$m, corresponding to the nuclear scale, is however strongly constrained by experiment. The role of the smearing scale and its interpretation are discussed in detail in reviews and more recent analyses, e.g.\ \cite{BassiGhirardi2003,CarlessoEtAl2022,Figurato:2024tpo}. As discussed in section 5, the stringy model naturally includes a built-in UV regulator (at a scale much larger than the string distance).

\section{DP model and experiment}
\label{sec:dp_constraints}

A collapse model is judged by two distinct criteria: (i) \emph{effectiveness} --- the collapse must be rapid enough to account for the emergence of definite outcomes in genuinely macroscopic, pointer-like superpositions; and (ii) \emph{consistency} --- the same dynamics must remain compatible with laboratory and astrophysical bounds on its unavoidable side effects, such as diffusion/heating and, in some versions, spontaneous radiation. In this section, we briefly review how these criteria apply to the standard DP model. 

\subsection{Effectiveness: ``fast enough'' collapse}
\label{sec:dp_fast}

The DP master equation implies a collapse rate controlled by the gravitational self energy of the difference between the two mass distributions, $\Gamma_{\rm coll}\sim \Delta E_G/\hbar$ (see \cref{eq:Penrose_tau,eq:Penrose_DeltaEG}). For order of magnitude estimates, and for superpositions whose branch separation exceeds the DP smearing scale ($\Delta x\gg R_0$), one finds parametrically
\begin{equation}
\Gamma_{\rm coll}\ \sim\ \frac{G M^2}{\hbar\,R_0}\; {\cal G}(\text{geometry},\Delta x),
\label{eq:DP_coll_rate_scaling}
\end{equation}
where ${\cal G}$ is an order one (or at most mildly logarithmic) geometric factor that depends on the shape of the object and on the branch separation. Thus, smaller values of $R_0$ lead to faster collapse.

What counts as ``fast enough'' is not determined by experiment alone; rather, it is a phenomenological requirement that any objective collapse proposal must satisfy. A careful recent analysis of the DP model's effectiveness is given in \cite{Figurato:2024tpo}, where the predicted DP collapse times for macroscopic objects are compared with physically motivated timescales for the formation of a stable classical record.

\subsection{Consistency with experiments}
\label{sec:dp_expt}

To the best of our knowledge, two classes of experiments are particularly relevant for the standard DP model, as well as for the stringy version discussed below.

{\it (1) High frequency bound:}
The standard white/Markovian Diósi–Penrose model is strongly constrained by spontaneous-radiation searches: an important dedicated bound was obtained at Gran Sasso \cite{Donadi2021}, later germanium-based analyses with the MAJORANA Demonstrator \cite{ArnquistEtAl2022} strengthened this line of constraint, and the current strongest bound now comes from XENONnT \cite{XENON:2025plg} 
\be\label{R0}
R_0> 4.4  \times 10^{-10}  ~~~~~~~~(95\%).
\ee
This bounds, however, do not directly apply to a colored-noise realization whose temporal spectrum is strongly suppressed at X-ray frequencies \cite{AdlerBassi2007,CarlessoFerialdiBassi2018}.

{\em (2) Low frequency bounds:}
Constraints based on low frequency force noise probe the mHz band and do not rely on whiteness at keV--MeV frequencies. In particular, LISA Pathfinder was an ESA technology mission designed to demonstrate that two test masses inside a single spacecraft could be maintained in near perfect free fall, as required for space based gravitational wave detection. It measured the residual differential acceleration, or equivalently relative displacement noise, between the masses at mHz frequencies using a laser interferometric readout together with a drag free control system that steered the spacecraft to follow one of the masses. Such experiments provide weaker, but more robust, bounds on the same underlying diffusion strength \cite{Helou2017LPF}.

The key point is that, in the standard DP model, the same UV scale $R_0$ controls both collapse effectiveness and the size of the experimentally constrained side effects. Parametrically,
\begin{equation}
\Gamma_{\rm coll}\ \propto\ \frac{1}{R_0},
\qquad\qquad
D_{\rm DP}\ \propto\ \frac{G}{R_0^{3}},
\label{eq:DP_two_scalings}
\end{equation}
where $D_{\rm DP}$ denotes a representative momentum diffusion or heating strength constrained by experiment. Increasing $R_0$ suppresses the experimentally dangerous diffusion/heating rather efficiently ($\propto R_0^{-3}$), but at the same time it also suppresses collapse ($\propto R_0^{-1}$).

As emphasized in \cite{Figurato:2024tpo} (see also the general review \cite{BassiDoratoUlbricht2023}), once one imposes the experimental lower bound (\ref{R0}) on $R_0$ (or comparable non interferometric constraints), the remaining allowed parameter region tends to correspond to collapse that is too slow to provide a compelling objective collapse solution to the measurement problem for mesoscopic to macroscopic superpositions. This is the sense in which the standard white noise DP model is often said to be ``strongly constrained'': the parameter region that is clearly compatible with current data is typically not phenomenologically attractive as a collapse theory.

\section{A point particle toy model}
\label{sec:lin_view}

In this section, we begin to address the question of whether there exists a physical system, perhaps a non-standard cosmological fluid, capable of giving rise to DP-like behavior. To this end, we introduce a toy model formulated in a point-particle framework. This model, however, suffers from several obvious shortcomings. In the next section, we argue that string theory, in a suitable cosmological setting, provides a UV completion of this toy model that is free of the problems inherent in the point-particle description.

A natural first attempt is to consider a bath of uncorrelated particles containing both positive energy, $+E$, and negative energy, $-E$, quanta, with equal densities so that $\langle T_{\mu\nu}\rangle=0$ and hence no effect on the homogeneous evolution of the universe. This idea fails for two reasons, one technical and one conceptual. The technical problem is that this monopole shot-noise model gives (see Appendix A)
\begin{equation}
\label{eq:Phi_shot}
\langle \Phi\Phi\rangle \propto \frac{1}{k^4},
\end{equation}
which is redder than the DP scaling, $1/k^2$. The conceptual problem is even more serious: the proposal suffers from a fatal instability. If the positive and negative energy quanta are not microscopically correlated, but merely happen to have equal densities on large scales, then the total energy is unbounded from below. In such a scenario, arbitrarily large numbers of negative energy quanta could be produced.

These two problems are, in fact closely related. To cure the conceptual instability, the negative energy particle must be linked microscopically to the positive energy particle. One is then naturally led to excitations with vanishing monopole gravitational charge (energy), so that the leading gravitational effect is dipolar. Quite generally, this leads to a DP-like spectrum,
\be
\label{eq:toy_DP_spectrum}
\langle \Phi\Phi\rangle \propto \frac{1}{k^2},
\ee
as can be illustrated in a simple toy model in which the fundamental excitation is a dipole: a particle with positive energy, $+E$, located at $\mathbf x-\frac{\mathbf d}{2}$, and a particle with negative energy, $-E$, located at $\mathbf x+\frac{\mathbf d}{2}$, so that
\begin{equation}
\label{eq:dipole_T00_real}
T_{00}(\mathbf x)=E\,\delta^{(3)}\!\left(\mathbf x-\frac{\mathbf d}{2}\right)
- E\,\delta^{(3)}\!\left(\mathbf x+\frac{\mathbf d}{2}\right).
\end{equation}
The Fourier transform of \cref{eq:dipole_T00_real} is
\begin{equation}
\label{eq:dipole_T00_k}
T_{00}(\mathbf k)=E\left(e^{-i\mathbf k\cdot \mathbf d/2}-e^{+i\mathbf k\cdot \mathbf d/2}\right)
=-2iE\,\sin(\mathbf k\cdot\mathbf d/2),
\end{equation}
which, at small $k$, gives
\begin{equation}
\label{eq:dipole_smallk}
\lvert T_{00}(\mathbf k)\rvert^2 \approx E^2\,(\mathbf k\cdot \mathbf d)^2 \propto k^2.
\end{equation}
This in turn implies \cref{eq:toy_DP_spectrum}; see Appendix B for details.

This static dipole model reproduces the spatial DP-like scaling, $1/k^2$. However, it does not yield, even approximately, white noise in time. Instead of $\sim \delta(\omega+\omega')$, one obtains $\sim \delta(\omega)\delta(\omega')$, namely a static disorder that does not induce wavefunction collapse. The following modification of the toy model remedies this. Rather than considering static dipoles with a fixed density, we consider dynamical dipoles characterized by a timescale $\tau_{\rm dip}$ and a creation rate (events per four volume) $\Gamma_{\rm dip}$. The simplest example is an instant dipole: a dipole that is created instantaneously and persists for a lifetime $\tau_{\rm dip}$. For simplicity, we consider a top-hat model in which the dipole is created sharply at full size $d$ at time $t_0$ and disappears just as sharply at $t_0+\tau_{\rm dip}$. This model is clearly unrealistic, since each particle entails a local violation of energy conservation both at production and at annihilation. 
A further limitation of this model is that it does not specify a mechanism for the production of these dipoles.

Nevertheless, we set these issues aside (as they do not arise in the stringy model discussed in the next section) and analyze the effect generated by this toy setup. The Fourier transform in time is simple and gives
$
%|\tilde \chi(\omega)|^2 = 
4\,\sin^2(\omega\tau_{\rm dip}/2)/\omega^2,
$
which means that the noise is approximately white in time in the band $\omega\ll 1/\tau_{\rm dip}$, and is suppressed for $\omega\gg 1/\tau_{\rm dip}$. For $\omega\ll 1/\tau_{\rm dip}$ the power spectrum is
\begin{equation}
P_{\Phi}(\mathbf{k},\omega)
\sim
G^2\;\Gamma_{\rm dip}\;E^2 d^2\;\tau_{\rm dip}^2\;\frac{1}{k^2},
\label{eq:Pphi_plateau}
\end{equation}
which implies a DP-like model with
\be
G\to G^2\;\Gamma_{\rm dip}\;E^2 d^2\;\tau_{\rm dip}^2,
\ee
with $\hbar=1$. A useful feature of this toy model is that it comes with a built-in UV cutoff, namely the dipole size. Thus, roughly,
\be
R_0 \to d,
\ee
which is a free parameter of the model rather than something tied to the nuclear scale.

A closely related toy model, which more closely approximates the stringy setup discussed in the next section, is the growing dipole model. In this case, the dipole separation grows ballistically after creation, say at $t=0$, for some time $\tau_{dip}$
\begin{equation}
d(\tau)= v \tau,\qquad 0<\tau<\tau_{\rm dip}.
\end{equation} 
This model is also unrealistic for two reasons: (i) There is no creation mechanism, and as a result, the production rate, $\Gamma_{\rm dip}$ is not known. (ii) Each particle violates energy and momentum conservation upon annihilation.\footnote{Unlike the previous toy model, here, the second issue can be resolved via interactions between nearby dipoles.}

\begin{figure}[t!]
\centering
\includegraphics[width=16cm]{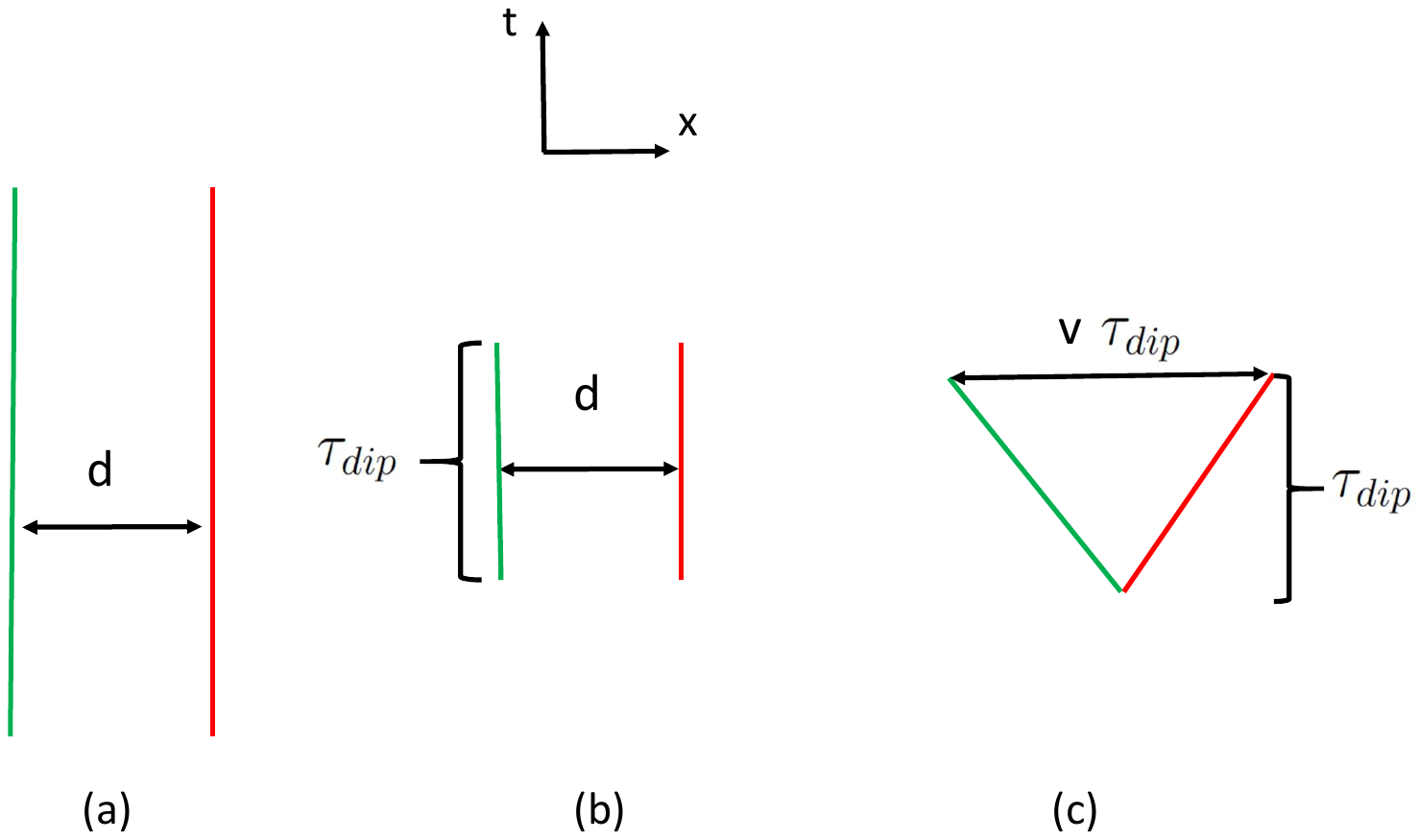}
\vspace{-2cm}
\caption{Toy dipole models. The green line represents the positive energy particle, and the red line the negative energy particle. Panel (a) reproduces the DP spatial dependence, $1/k^2$, but does not yield approximately white noise in time. Panels (b) and (c) do. Panel (c) provides a good approximation to the stringy model discussed in the next section.}
\label{fig:toy_models}
\end{figure}

Setting these issues aside once more, one finds (see Appendix C for details)
\begin{equation}
P_{\Phi}(\mathbf{k},\omega)\ \simeq\
\frac{(4\pi G)^2}{12}\,
\Gamma_{\rm dip}\,\,E^2 v^2\;
\tau_{\rm dip}^4\;\frac{1}{k^2},
\label{bb}
\end{equation}
for $\omega\ll 1/\tau_{\rm dip}$.
Matching to the standard DP scaling, we may identify the corresponding effective DP parameters of the growing dipole toy model as
\be
G\to G^2\;\Gamma_{\rm dip}\;E^2 v^2\;\tau_{\rm dip}^4,
\qquad\mbox{and}\qquad
R_0\to v\tau_{\rm dip},
\ee
as effective parameters.

\section{The stringy model}
\label{sec:IFS}

A different fundamental question, seemingly unrelated to the measurement problem in quantum mechanics, concerns the origin of cosmic acceleration. In string theory, this issue has been the subject of ongoing debate ever since it was discovered, almost thirty years ago, that the universe is accelerating \cite{Riess1998,Perlmutter1999}. Recently, \cite{Itzhaki:2024pok} proposed a novel mechanism in which the present acceleration of the universe is driven by Instant Folded Strings (IFSs) and their decay products. Although IFSs are non-standard strings with unconventional properties \cite{Attali:2018goq}, it is not straightforward to identify a simple cosmological signature that cleanly distinguishes IFS-driven acceleration from more conventional scenarios. The reason is that, on cosmological scales, IFSs and their decay products collectively behave as an effectively ordinary cosmological fluid\footnote{I thank Ely Kovetz, Uri Peleg and Paul Steinhardt for discussions on this point.}.

The purpose of this section is to show that, in a certain region of parameter space, IFS-driven acceleration leaves a distinctive imprint by giving rise to a DP-like model. We begin by reviewing the main properties of IFSs and their decay products. We then show that, at distances large compared to the string scale, their dynamics can be approximated by the growing dipole toy model introduced in the previous section (without its inconsistency). In other words, in the setup considered in \cite{Itzhaki:2024pok}, string theory appears to provide a UV completion of the Diósi–Penrose model.

An IFS is an unusual closed string that is nucleated \emph{classically} when the string coupling grows with time. Shortly after nucleation, the string expands asymptotically at the speed of light. The energy-momentum tensor associated with an IFS created at $(t_0, x_0, y_0, z_0)$ and growing in the $x$-direction is \cite{Attali:2018goq}
\be
\label{eq:ifs_tmunu}
\begin{aligned}
&T_{uu}=-\frac{v-v_0}{2\pi\alpha'}\Theta(v-v_0)\delta(u-u_0)\delta^{(2)}_\perp(y-y_0,z-z_0), \\
&T_{vv}=-\frac{u-u_0}{2\pi\alpha'}\Theta(u-u_0)\delta(v-v_0)\delta^{(2)}_\perp(y-y_0,z-z_0), \\
&T_{uv}=\frac{1}{2\pi\alpha'}\Theta(u-u_0)\Theta(v-v_0)\delta^{(2)}_\perp(y-y_0,z-z_0),
\end{aligned}
\ee
where the light-cone variables are defined by $v \equiv t + x$ and $u \equiv t - x$, and
\[
\delta^{(2)}_\perp(y - y_0, z - z_0) \equiv \delta(y - y_0)\delta(z - z_0).
\]
The component $T_{uv}$ describes the positive energy density in the bulk of the string, arising from the string tension. This positive contribution is exactly canceled by the negative null energy at the folds, described by $T_{uu}$ and $T_{vv}$. This cancellation is not accidental: it follows from the fact that the energy vanishes before the string is created, $t<t_0$, and therefore continues to vanish for $t>t_0$\footnote{This statement is exact in a timelike linear-dilaton background. In the setup of interest here \cite{Itzhaki:2024pok}, it receives a small correction due to the cosmological evolution. We will neglect this correction in the present discussion, since it amounts schematically to $1/k^2 \to 1/k^2 +H^2/k^4$ and thus becomes important only on cosmological scales.}. The upshot is that the monopole gravitational charge vanishes. Parity dictates that the dipole moment vanishes. The quadrupole is therefore the first non-vanishing moment. It is given by
\be
Q(t)=-\frac{4}{3\pi\alpha'}\,(t-t_0)^3\,\Theta(t-t_0),
\ee
which implies that
\begin{equation}
P_{\rm GW}(t) \sim
\frac{G}{\alpha'^2}\;\Theta(t-t_0),
\end{equation}
is the total gravitational wave luminosity. Thus, an IFS radiates at constant power. Since $G\sim g_s^2 $, this radiation is, as expected, small at weak coupling. Nevertheless, it is clear that, no matter how small $g_s$ is, this process cannot continue indefinitely. 

\begin{figure}[t!]
\centering
\includegraphics[width=16cm]{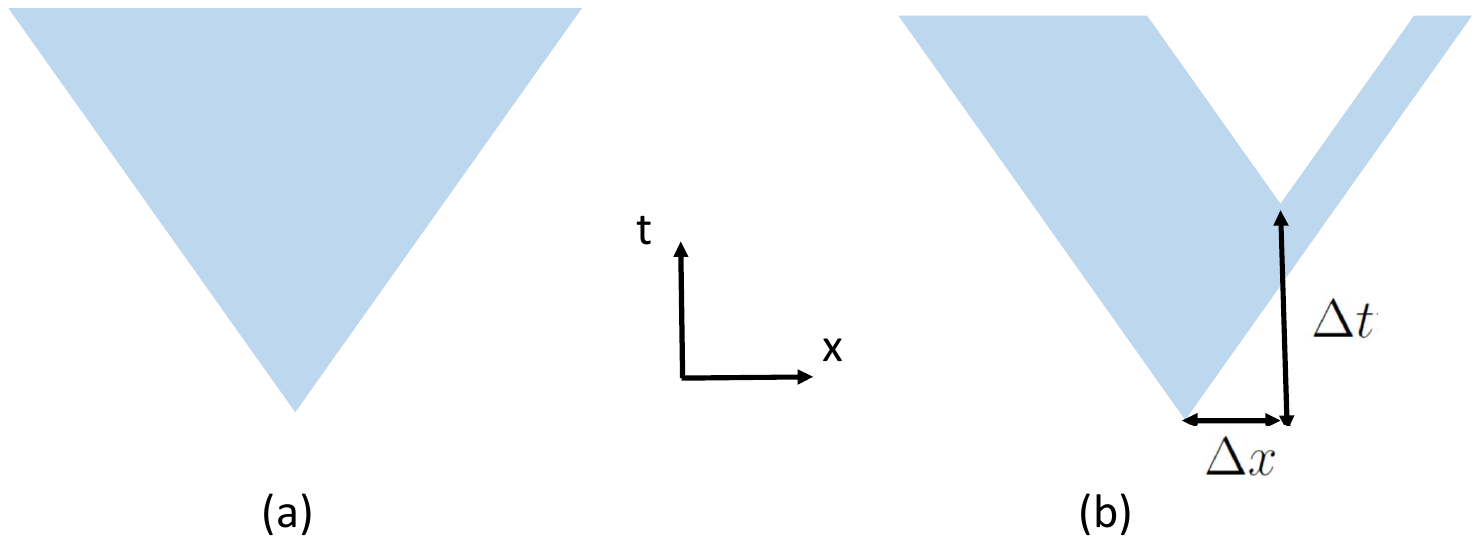}
\vspace{-2cm}
\caption{(a) An instant folded string is created classically at a point and subsequently expands at the speed of light, with its total energy remaining zero. (b) Quantum mechanically, an instant folded string splits. Because its total energy is zero, the decay produces an energy-EPR state that behaves as an expanding dipole originating at the splitting point. In this example, the left-moving string carries positive energy, while the right-moving string carries negative energy, so the net energy remains zero.}
\label{fig:ifs}
\end{figure}

Indeed, at finite $g_s$, an IFS, like any other string, can break. Because the total energy of an IFS vanishes, something rather unusual happens when it splits: an energy EPR state is formed \cite{ItzhakiPeleg2024}. As we now explain, this energy EPR state is well approximated by the growing dipole toy model discussed in the previous section. Suppose that, a time $\Delta t$ after its creation, the IFS splits into two, and that the splitting point lies a distance $\Delta x$ away from the point at which the IFS was created (see \cref{fig:ifs}). The result is two closed folded strings with energies
\be
E_1=-E_2=\frac{\Delta x}{2\pi \alpha'}.
\ee
The splitting of an instant string is a local process that does not depend on the position of the splitting point, as long as it is away from the fold. Hence, the wavefunction of the two closed strings takes the form
\begin{equation}
\ket{\Psi(\Delta t)}
\sim
\int_{-\frac{\Delta t}{2\pi\alpha'}}^{\frac{\Delta t}{2\pi\alpha'}}
dE_1\;
\ket{E_1=\Delta x/2\pi\alpha'}
\otimes
\ket{E_2=-E_1},
\end{equation}
which describes an energy EPR state \cite{ItzhakiPeleg2024}. 

We expect the typical splitting time to scale as 
$
\tau_{\rm IFS}\sim l_s/g_s,
$
which means, using \(\Delta x \sim \tau_{\rm IFS}\) and \(\tau_{\rm IFS}\sim \ell_s/g_s\), that the typical splitting energy is
\begin{equation}
E_{\rm typ}
\sim \frac{\Delta x}{2\pi \alpha'}
\sim \frac{\tau_{\rm IFS}}{\alpha'}
\sim \frac{\ell_s/g_s}{\ell_s^2}
\sim \frac{1}{g_s \ell_s}
\sim M_{\rm Pl}.
\end{equation}

As is clear from \cref{fig:ifs}, the right-moving (left-moving) string continues to move to the right (left) at the speed of light. As a result, after the splitting, the quadrupole grows only linearly with time, and the gravitational wave emission ceases. More importantly, a dipole moment is generated:
\be
D(t)=2E_1\,\tilde t\,\Theta(\tilde t),
\qquad
\tilde t \equiv t-\Delta t-t_0 .
\label{eq:dipole-growth}
\ee
This dipole grows linearly with time, just as in the growing dipole discussed at the end of the previous section, with $v=1$. A more precise statement is that, because the post splitting state is an energy EPR state, the dipole moment has vanishing expectation value (due to the integral over $E_1$), but nonzero variance:
\be
\bra{\Psi(t)}D\ket{\Psi(t)}=0, \qquad
\bra{\Psi(t)}D^2\ket{\Psi(t)} \sim M_{\rm Planck}^2 \tilde{t}^2.
\ee

After roughly $l_s/g_s$ another splitting occurs. As more and more splittings and radiation take place, the dipole growth effectively stops, and it is therefore natural to expect that
\be
\label{eq:C_timescale}
\tau_{\rm dip}=C \frac{l_s}{g_s},
\ee
with $C\gg 1$. 
%As we discuss in the next section, $C$ may be of phenomenological relevance, and it is therefore important to know how large it is. 
%The expectation that $C$ should be large follows from energy-momentum conservation: the energy and momentum of, say, the left-moving string are conserved, so its decay products continue, as a whole, to move to the left. The same argument applies to the right-moving string, which means that the dipole moment tends to keep growing, despite the extra splittings. Hence numerous interactions are required to suppress the dipole growth. 
%In a sense, the bug of the point-particle toy model, that each particle violates energy and momentum conservation upon
%annihilation, turns into a feature in the stringy model, that $C\gg 1$. Unfortunately, it is difficult to estimate how large this feature actually is, owing both to the large number of available decay channels and to the complexity of the exact CFT description of an IFS \cite{Hashimoto:2022dro}. In fact, depending on the value of $g_s$ it is possible that the channel that suppresses the most the growth of the dipole involves interactions with nearby energy-EPR states. 

As we discuss in the next section, $C$ may be phenomenologically relevant, and it is therefore important to know how large it can be. The expectation that $C$ should be large follows from energy and momentum conservation: the energy and momentum of, say, the left-moving string are conserved, so its decay products, taken as a whole, continue to move to the left. The same reasoning applies to the right-moving string. As a result, the dipole moment tends to keep growing despite the additional splittings, and many interactions are therefore required before this growth is significantly suppressed.

In this sense, what appeared as a bug of the point-particle toy model - namely, that each annihilation event violates, locally, energy and momentum conservation - turns into a feature in the stringy model, suggesting that $C\gg 1$. Unfortunately, it is hard to estimate how large $C$ actually is, both because it is difficult to pinpoint which of the many channels dominants the dipole suppression and because of the complexity of the exact CFT description of an IFS \cite{Hashimoto:2022dro}. Moreover, depending on the value of $g_s$, it is possible that the channel that suppresses the dipole growth most efficiently involves interactions between nearby energy-EPR states. In such a case $C$ should depend also on the production rate of the IFSs as well and on $H_0$.

In short, within the setup discussed in \cite{Itzhaki:2024pok}, $C$ is in principle calculable. Unfortunately, at present, we are not in a position even to estimate it. We can, however, estimate the IFSs production rate.
If the current cosmic acceleration is indeed induced by IFSs, then, up to an order one constant, we have \cite{Itzhaki:2024pok}
\be
\Gamma_{\rm IFS}\sim 10^{-120}
\ee
in Planck units. Combining this with \ref{eq:C_timescale} and (\ref{bb}), we obtain a DP-like model with
\begin{equation}
G_{\rm eff}=G\frac{10^{-120}}{g_0^{8}},
\qquad
R_{0,{\rm eff}}=\frac{\ell_{\rm Pl}}{g_0^{2}}, \qquad \mbox{with} \qquad g_0\equiv g_s/\sqrt{C}
\label{eq:IFS_Geff_R0_gbar}
\end{equation}
as effective parameters.

\section{Comparison to experiment}
\label{sec:ifs_dp_constraints}

We now apply the same two criteria used in \cref{sec:dp_constraints} - (i) collapse must be sufficiently fast, and (ii) the associated noise must be consistent with current experimental bounds - to the IFS induced DP-like model. As we will see, a key qualitative difference from the standard DP model is that the IFS induced DP-model is colored in time.

\subsection{Effectiveness: ``fast enough'' collapse}
\label{sec:ifs_fast}

For superpositions with separation $\Delta x\gg R_{0,{\rm eff}}$, the DP scaling estimate applies with $G\to G_{\rm eff}$ and $R_0\to R_{0,{\rm eff}}$:
\begin{equation}
\Gamma_{\rm coll}\ \sim\ \frac{G_{\rm eff}M^2}{\hbar\,R_{0,{\rm eff}}}
=
\frac{G M^2}{\hbar\,\ell_{\rm Pl}}\;10^{-120}\,g_0^{-6},
\label{eq:IFS_Gamma_coll}
\end{equation}
up to geometric factors of order one. Requiring collapse within a target time $T$ (that is, $\Gamma_{\rm coll}\gtrsim 1/T$) yields
\begin{equation}
g_0 \ \lesssim\
1.26\times 10^{-15}\,
\left(\frac{M}{10^{-14}\ {\rm kg}}\right)^{1/3}
\left(\frac{T}{1\ {\rm s}}\right)^{1/6}.
\label{eq:IFS_gbar_max}
\end{equation}
This gives a concrete upper bound on $g_0$ once one specifies a phenomenological target $(M,T)$ for what counts as ``fast enough'' collapse. For orientation, the following benchmark choices give:
\begin{center}
\renewcommand{\arraystretch}{1.2}
\begin{tabular}{@{}lccc@{}}
\toprule
Target (phenomenological) & $M$ [kg] & $T$ [s] & $g_0^{\max}$ \\
\midrule
Micron dust grain collapses in $1$ ms & $10^{-15}$ & $10^{-3}$ & $1.85\times 10^{-16}$ \\
Nanosphere collapses within $1$ s & $10^{-17}$ & $1$ & $1.26\times 10^{-16}$ \\
Mesoscopic ``pointer'' collapses within $1$ s & $10^{-14}$ & $1$ & $1.26\times 10^{-15}$ \\
Nanogram pointer collapses in $1$ ms & $10^{-12}$ & $10^{-3}$ & $1.85\times 10^{-15}$ \\
Microgram pointer collapses in $1$ $\mu$s & $10^{-9}$ & $10^{-6}$ & $5.85\times 10^{-15}$ \\
\bottomrule
\end{tabular}
\end{center}
The mild scaling $g_0^{\max}\propto M^{1/3}T^{1/6}$ implies that even large changes in $(M,T)$ shift the bound only modestly. Hence,
\be
g_0 \ \lesssim\ 10^{-15}
\ee
is a reasonable order of magnitude estimate for effectiveness.

\subsection{Consistency with experiments}
\label{sec:ifs_expt}

The strongest non-interferometric constraints typically bound diffusion and heating effects which, in the white noise DP model, scale as $G/R_0^3$. In the IFS-induced DP-like model, one finds
\begin{equation}
\frac{G_{\rm eff}}{R_{0,{\rm eff}}^{3}}
=
\frac{G}{\ell_{\rm Pl}^{3}}\,10^{-120}\, g_0^{-2}.
\label{eq:IFS_diff_strength}
\end{equation}
Experiments, therefore, place lower bounds on $g_0$, since values of $g_0$ that are too small would overproduce diffusion and heating.

To determine which experimental bounds are relevant, one must compare the probe frequency with the correlation frequency cutoff of the colored noise. The relevant temporal cutoff is
\begin{equation}
\omega_c
\sim
\frac{c}{R_{0,\rm eff}}
\sim
\frac{c\,g_0^2}{\ell_{\rm Pl}},
\label{eq:omega-c-SI}
\end{equation}
where we have restored the speed of light, $c$. Thus, whiteness up to a probe frequency $\omega$ requires $\omega \ll \omega_c$.

If one assumes white noise in time, XENONnT \cite{XENON:2025plg}  gives the strongest bound (\ref{R0}). However, the noise in the IFS-induced DP-like model is not white. Whiteness up to $\omega_{\rm X}\sim 10^{19}\,{\rm s^{-1}}$ would require $g_0\gtrsim 10^{-12}$. In the collapse motivated window of \cref{eq:IFS_gbar_max}, one typically has $g_0\sim 10^{-16}$--$10^{-15}$, and therefore $\omega_c\sim 10^{11}$--$10^{13}\,{\rm s^{-1}}$, far below X-ray frequencies. Spontaneous emission bounds are therefore expected to be parametrically suppressed.

By contrast, robust low-frequency bounds, such as those from LISA Pathfinder, do not rely on high-frequency whiteness and imply only that $g_0\gtrsim 10^{-27}$. This bound is most likely negligible compared with the fact that $g_0\sim 10^{-15}$ places the string scale roughly at $10\,\sqrt{C}\,{\rm TeV}$. Unless $C$ is exponentially large, the range $g_0\lesssim 10^{-27}$ is already ruled out by LHC. This highlights the importance of estimating $C$: if it is not excessively large, then, within the setup considered here, string theory may lie within the reach of future accelerators.

In summary, mainly because, unlike the original DP model, the stringy version is colored in time, it appears to be less constrained by existing experimental bounds. A secondary reason is that the relevant parametric scalings are also different,
\begin{equation}
\Gamma_{\rm coll}\ \propto\ g_0^{-6},
\qquad\qquad
D\propto\ \frac{G_{\rm eff}}{R_{0,{\rm eff}}^{3}}\ \propto\ g_0^{-2},
\label{eq:IFS_two_scalings}
\end{equation}
compared to \ref{eq:DP_two_scalings}. As a result, one can make a collapse fast while the diffusion and heating strength grow more slowly.

\section{Discussion}

In this work, we have shown that if IFSs and their decay products are responsible for the present cosmic acceleration \cite{Itzhaki:2024pok}, then they also generate a two-point correlator for the Newtonian potential of the DP form. It is worth noting that, in \cite{Itzhaki:2018glf,Itzhaki:2023hwm}, it was conjectured that other instant objects, such as instant folded D-branes, may also arise and influence a wider class of time-dependent backgrounds in string theory. This raises the possibility that DP-related backgrounds are a more generic feature of string theory.

The central conceptual issue, which we wish to address now, is how this result should be interpreted. Does the appearance of a DP-like kernel point to a genuine, fundamental mechanism of wavefunction collapse, or does it instead reflect effective stochasticity generated by coarse-graining over unobserved degrees of freedom? IFSs nucleation is modeled as a classical stochastic process, and this in turn induces an apparently stochastic contribution to the Newtonian potential $\Phi$. The key question is whether this stochasticity should be regarded as fundamental, namely as an intrinsic random component of $\Phi$ that cannot be purified 
and hence drives genuinely non-unitary evolution, or as effective, 
while the full system, schematically, ``matter + gravity + IFS'', continues to evolve unitarily.

If the latter interpretation is correct, then the DP-like structure of the correlator does not by itself imply a modification of quantum mechanics. Rather, it characterizes the effective noise kernel that governs the reduced dynamics of matter fields coupled to a gravitational environment populated by IFSs. In that case, the appearance of a DP kernel is analogous to other situations in which a subsystem acquires a stochastic description only because part of the full quantum state has been coarse-grained over.

Interestingly, the answer appears to depend sensitively on the global structure of spacetime. In asymptotically flat or asymptotically 
AdS settings, the apparent stochasticity induced by IFSs is expected to be purifiable, rather than corresponding to genuine wavefunction collapse.
%in a larger Hilbert space. 
A useful example that illustrate this is a stack of black NS5-branes \cite{Strominger:1990et} in asymptotically Minkowski spacetime. In such a background, IFSs are produced behind the horizon \cite{Itzhaki:2018glf}. If one keeps track only of coarse-grained quantities, such as their density, their contribution to $\Phi$ appears stochastic. Nevertheless, this stochasticity need not be fundamental: it is natural to expect that the exact quantum state of the IFS sector is determined by the quantum state of the matter that formed the black NS5-brane in the first place. In this sense, the DP-like kernel should be interpreted as an effective noise correlator. It signals decoherence induced by the IFS sector, but it does {\it not} cause a collapse of the wavefunction.

The situation is qualitatively different in cosmology. In a de~Sitter-like universe, one has access only to observables within a causal patch, and the very definition of global observables is considerably subtler than in asymptotically flat or $AdS$ space \cite{Witten2001}. In such circumstances, recoherence may be impossible not merely in practice but operationally, since the degrees of freedom required to purify the state are never accessible within a fixed Hubble patch. The distinction between fundamental and effective stochasticity may then become empirically inaccessible.

This point is sharpened further in the recently proposed IFS-induced cyclic cosmology \cite{Itzhaki:2025gdv}. 
There, entropy, black holes, and more generally any environmental ``records'' produced in one cycle are rendered locally irrelevant in later cycles, since the prolonged accelerated expansion and the net growth of the scale factor from cycle to cycle exponentially dilute them and, in the newer formulations \cite{SteinhardtTurok2002PRD,IjjasSteinhardt2019NewCyclic,IjjasSteinhardt2022Entropy}, can push them beyond the observable horizon. As a result, the information needed to purify the state within any fixed Hubble volume is dispersed over an enormous region in subsequent cycles. The cosmological evolution, therefore, implements a dynamical coarse-graining that renders the IFS induced stochastic contribution to $\Phi$ irreducible during the dark energy epoch.

From this perspective, the DP-like kernel acquires a stronger interpretational significance. Since the full ``matter + gravity + IFS'' state is not accessible, the effective stochastic description appears to be operationally indistinguishable from a genuinely stochastic one. It is therefore natural to adopt a collapse-like interpretation.

The sensitivity to the global structure of spacetime can be sharpened by a more formal question: can one define, even in principle, the Hilbert space in a quantum theory that includes gravity without detailed microscopic input? As emphasized by Witten \cite{Witten2001}, the answer appears to depend crucially on the background geometry. The clearest case is AdS, where the asymptotic behavior at spatial infinity determines the Hilbert space and observables, providing the foundation for the AdS/CFT correspondence. In asymptotically Minkowski space the S-matrix provides a sharp observable. 
However, it remains unclear whether it is possible to go beyond this by defining additional gauge-invariant observables, analogous to Wilson lines in AdS/CFT \cite{Maldacena:1998im,Rey:1998ik}, or by formulating the holographic principle \cite{tHooft:1993dmi,Susskind:1994vu} in a precise way (for a recent approach see \cite{PasterskiShaoStrominger2017,MeltonSharmaStromingerWang2024}).

The situation is much more restrictive in cosmology. In de Sitter-like spacetime, the only available asymptotic regions lie in the far past and far future, so one is naturally led to define ``meta-observables'' that have access to those asymptotic structures.  By construction, however, such quantities are inaccessible to any observer confined to a single causal patch. In a Big-Bang cosmology that becomes asymptotically de Sitter in the far future (e.g. $\Lambda$CDM), the future asymptotic region is the only structure from which one might attempt to construct such meta-observables. In the IFS-induced cyclic bouncing universe \cite{Itzhaki:2025gdv}, the dark energy epoch is transient, and there is no asymptotic future de Sitter boundary either; from this viewpoint, it is not clear that meta-observables exist at all \cite{Witten2001}, and it is not clear how to define, even in principle, a Hilbert space. This provides an additional sense in which the \cite{Itzhaki:2025gdv} setting is naturally ``patch-based'' and strengthens the motivation to treat the IFS-induced DP-like kernel as an objective stochastic ingredient during the epoch where it operates.

In summary, IFS dynamics provides a concrete microscopic mechanism for generating a DP-type noise kernel. Whether this kernel should be interpreted as supporting fundamental collapse, or instead as a manifestation of effective decoherence, appears to depend not only on the underlying microphysics but also, crucially, on the global structure of spacetime. 
In this way, the emergence of a DP-like correlator suggests a nontrivial link between infrared gravitational dynamics, horizon structure, and the foundations of quantum mechanics. We find it particularly appealing that this connection appears to be realized through string theory.

\vspace{6mm}
\noindent
{\bf Acknowledgments}

\noindent
I thank Eliahu Cohen for helpful comments.
Work supported in part by the ISF through grant number 256/22.

\appendix
\section{Monopole shot noise }
\label{app:shotnoise_1_over_k4}

In this appendix we show that a bath of uncorrelated pointlike quanta with energies $+E$ and $-E$, with equal number densities so that
$\langle T_{\mu\nu}\rangle=0$, nevertheless produces Newtonian-potential fluctuations with the characteristic monopole shot-noise spectrum
$\langle \Phi\Phi\rangle\propto 1/k^{4}$.

We work in linearized gravity in the Newtonian regime, where the scalar potential $\Phi(\mathbf{x},t)$ is sourced by
the energy density $T_{00}$ via the Poisson equation:
\begin{equation}
\nabla^2 \Phi(\mathbf{x},t)=4\pi G\,\delta T_{00}(\mathbf{x},t),
\qquad
\delta T_{00}\equiv T_{00}-\langle T_{00}\rangle,
\label{eq:poisson_app}
\end{equation}
and in momentum space; 
\begin{equation}
\Phi(\mathbf{k},t)= -\,\frac{4\pi G}{k^2}\,\delta T_{00}(\mathbf{k},t).
\label{eq:phi_k_app}
\end{equation}
We model the bath as two independent Poisson point processes of number density $\rho$ each (so the total number density is $2\rho$):
\begin{equation}
T_{00}(\mathbf{x})=\sum_{a\in +} E\,\delta^{(3)}(\mathbf{x}-\mathbf{x}_a)
\;-\;\sum_{b\in -} E\,\delta^{(3)}(\mathbf{x}-\mathbf{y}_b),
\label{eq:T00_point_app}
\end{equation}
where the sets $\{\mathbf{x}_a\}$ and $\{\mathbf{y}_b\}$ are uncorrelated and homogeneous.
By construction,
\begin{equation}
\langle T_{00}\rangle= E\rho - E\rho = 0,
\end{equation}
so $\delta T_{00}=T_{00}$ in this model.

The Fourier transform of \eqref{eq:T00_point_app} is
\begin{equation}
T_{00}(\mathbf{k})=E\sum_{a\in +} e^{-i\mathbf{k}\cdot \mathbf{x}_a}
\;-\;E\sum_{b\in -} e^{-i\mathbf{k}\cdot \mathbf{y}_b}.
\label{eq:T00_k_app}
\end{equation}
For an uncorrelated Poisson process, the connected two-point function of the Fourier modes is $k$-independent\cite{PeeblesLSS1980} :
\begin{equation}
\left\langle T_{00}(\mathbf{k})\,T_{00}(\mathbf{k}')\right\rangle
=(2\pi)^3\delta^{(3)}(\mathbf{k}+\mathbf{k}')\,P_{T_{00}}(k),
\qquad
P_{T_{00}}(k)=\text{const}.
\label{eq:def_PT00}
\end{equation}
One can compute the constant by standard shot-noise reasoning: for each species,
\begin{equation}
\left\langle \sum_{a} e^{-i\mathbf{k}\cdot \mathbf{x}_a}\sum_{a'} e^{-i\mathbf{k}'\cdot \mathbf{x}_{a'}}\right\rangle_{\rm conn}
=(2\pi)^3\delta^{(3)}(\mathbf{k}+\mathbf{k}')\,\rho,
\end{equation}
and similarly for the negative-energy species. Cross-correlators between $+$ and $-$ vanish because the two point processes are independent.
Therefore,
\begin{equation}
P_{T_{00}}(k)
%E^2\rho + E^2\rho 
= 2\rho\,E^2,
\label{eq:PT00_value}
\end{equation}
is independent of $k$, and 
\begin{align}
\left\langle \Phi(\mathbf{k})\,\Phi(\mathbf{k}')\right\rangle
&=\left(\frac{4\pi G}{k^2}\right)\left(\frac{4\pi G}{k'^2}\right)
\left\langle T_{00}(\mathbf{k})\,T_{00}(\mathbf{k}')\right\rangle
\nonumber\\[4pt]
&=(2\pi)^3\delta^{(3)}(\mathbf{k}+\mathbf{k}')\,
(4\pi G)^2\,\frac{2\rho\,E^2}{k^4}.
\label{eq:PhiPhi_k4}
\end{align}
Thus the equal-time spatial spectrum of the Newtonian potential induced by an uncorrelated $\pm E$ bath is
\begin{equation}
P_{\Phi}(k)\;=\;(4\pi G)^2\,\frac{2\rho\,E^2}{k^4}.
\end{equation}

\section{Spectrum induced by Dipoles }

In this appendix, we show that if the fundamental excitations have vanishing \emph{monopole} gravitational charge event-by-event (so that the leading
fluctuation is a dipole), then the Newtonian potential fluctuations scale as $P_\Phi(k)\propto 1/k^2$ at small $k$.  This is the same spatial
scaling as in the DP kernel.

Consider an elementary dipole excitation centered at $\mathbf{X}$ consisting of a $+E$ quantum at $\mathbf{X}-\mathbf{d}/2$ and a $-E$ quantum at
$\mathbf{X}+\mathbf{d}/2$,
\begin{equation}
T_{00}^{\rm dip}(\mathbf{x})
=
E\,\delta^{(3)}\!\left(\mathbf{x}-\mathbf{X}+\frac{\mathbf{d}}{2}\right)
-
E\,\delta^{(3)}\!\left(\mathbf{x}-\mathbf{X}-\frac{\mathbf{d}}{2}\right).
\label{eq:T00_dip_real_app}
\end{equation}
The Fourier transform is
\begin{align}
T_{00}^{\rm dip}(\mathbf{k})
&=
\int d^3x\,e^{-i\mathbf{k}\cdot\mathbf{x}}\,T_{00}^{\rm dip}(\mathbf{x})
\nonumber\\
&=
E\left(e^{-i\mathbf{k}\cdot(\mathbf{X}-\mathbf{d}/2)}-e^{-i\mathbf{k}\cdot(\mathbf{X}+\mathbf{d}/2)}\right)
=
-2iE\,e^{-i\mathbf{k}\cdot\mathbf{X}}\sin(\mathbf{k}\cdot\mathbf{d}/2).
\label{eq:T00_dip_k_app}
\end{align}
For $k|\mathbf{d}|\ll 1$,
\begin{equation}
\sin(\mathbf{k}\cdot\mathbf{d}/2)=\frac{\mathbf{k}\cdot\mathbf{d}}{2}+{\cal O}(k^3),
\qquad\Rightarrow\qquad
T_{00}^{\rm dip}(\mathbf{k})\;=\;-iE\,e^{-i\mathbf{k}\cdot\mathbf{X}}\,(\mathbf{k}\cdot\mathbf{d})+{\cal O}(k^3).
\label{eq:T00_dip_smallk_app}
\end{equation}
Thus \emph{event-by-event} monopole cancellation implies that the leading Fourier amplitude is linear in $k$.

Now consider a homogeneous, isotropic bath of dipoles created at random positions $\mathbf{X}_n$, with number density $\rho_{\rm dip}$,
and with independent, isotropically distributed dipole orientations $\hat{\mathbf{d}}$ of fixed magnitude $|\mathbf{d}|=d$.
At equal time we may write
\begin{equation}
T_{00}(\mathbf{k})=\sum_{n} T_{00,n}^{\rm dip}(\mathbf{k}),
\end{equation}
with each term given by \eqref{eq:T00_dip_k_app}.  Because the dipoles are assumed uncorrelated (a Poisson point process in their \emph{centers}),
the connected power spectrum is shot-noise-like in the \emph{event centers} and is obtained by averaging the single-event power over orientations:
\begin{equation}
P_{T_{00}}(k)\;\simeq\;\rho_{\rm dip}\,\Big\langle \big|T_{00}^{\rm dip}(\mathbf{k})\big|^2 \Big\rangle_{\hat{\mathbf{d}}}.
\label{eq:PT00_dip_from_single_event_app}
\end{equation}
Using \eqref{eq:T00_dip_smallk_app},
\begin{equation}
\big|T_{00}^{\rm dip}(\mathbf{k})\big|^2
\simeq
E^2\,(\mathbf{k}\cdot\mathbf{d})^2
=
E^2\,k^2 d^2\,(\hat{\mathbf{k}}\cdot\hat{\mathbf{d}})^2.
\end{equation}
For isotropic orientations, $\langle(\hat{\mathbf{k}}\cdot\hat{\mathbf{d}})^2\rangle=1/3$, hence
\begin{equation}
P_{T_{00}}(k)\;\simeq\;\rho_{\rm dip}\,\frac{1}{3}\,E^2\,d^2\,k^2
\qquad (kd\ll 1).
\label{eq:PT00_dip_k2_app},
\end{equation}
which gives
\begin{equation}
P_\Phi(k)
\simeq\;
\frac{(4\pi G)^2}{3}\,\rho_{\rm dip}\,\,E^2\,d^2\,\frac{1}{k^2}
\qquad (kd\ll 1).
\label{eq:Pphi_dip_1_over_k2_app}
\end{equation}
Thus, once the monopole contribution cancels event-by-event and the leading fluctuations are dipolar, the Newtonian potential acquires the
DP-like spatial scaling $P_\Phi(k)\propto 1/k^2$ at long wavelengths.

\section{Growing dipole toy model}
\label{app:growing_dipole}

In this appendix, we show that a bath of growing dipole events produces a Newtonian-potential correlator with DP-like spatial scaling
$P_\Phi\propto 1/k^2$ at small $k$, and with an approximately white temporal spectrum for $\omega\ll \omega_c\sim 1/\tau_{\rm dip}$ that rolls off
for $\omega\gg \omega_c$.

We model $T_{00}$ as a superposition of independent, identical events occurring at random spacetime points $(t_n,\mathbf{X}_n)$ with
rate (events per 4-volume) $\Gamma_{\rm dip}$.  Each event is a dipole whose separation grows ballistically after creation:
\begin{equation}
\mathbf{d}(t)= v\,t\,\hat{\mathbf{n}},\qquad 0<t<\tau_{\rm dip},
\label{eq:d_growth}
\end{equation}
where $\hat{\mathbf{n}}$ is a random unit vector (isotropic distribution), $v\le 1$, and $\tau_{\rm dip}$ is the coherence/lifetime of the event.
The instantaneous energy density of a single event (centered at $\mathbf{X}$ and created at $t_0$) is taken to be
\begin{equation}
T_{00}^{\rm event}(\mathbf{x},t)=
E\,\delta^{(3)}\!\left(\mathbf{x}-\mathbf{X}-\frac{\mathbf{d}(t-t_0)}{2}\right)
-
E\,\delta^{(3)}\!\left(\mathbf{x}-\mathbf{X}+\frac{\mathbf{d}(t-t_0)}{2}\right),
\qquad 0<t-t_0<\tau_{\rm dip},
\label{eq:T00_event_real}
\end{equation}
and vanishing outside the interval $(t_0,t_0+\tau_{\rm dip})$. Fourier transforming \eqref{eq:T00_event_real} in space gives, for $0<t<\tau_{\rm dip}$,
\begin{equation}
T_{00}^{\rm event}(\mathbf{k},t)=
-2iE\,e^{-i\mathbf{k}\cdot\mathbf{X}}\sin\!\Big(\frac{\mathbf{k}\cdot\mathbf{d}(t)}{2}\Big).
\label{eq:T00_event_k_t}
\end{equation}
For $k\,d(t)\ll 1$ (long wavelengths compared to the dipole size at that time),
\begin{equation}
T_{00}^{\rm event}(\mathbf{k},t)\simeq
-iE\,e^{-i\mathbf{k}\cdot\mathbf{X}}\,(\mathbf{k}\cdot\mathbf{d}(t))
=
-iE\,e^{-i\mathbf{k}\cdot\mathbf{X}}\,(v t)\,(\mathbf{k}\cdot\hat{\mathbf{n}}).
\label{eq:T00_event_smallk}
\end{equation}
Thus the event amplitude is linear in both $k$ and $t$.

Fourier transforming in time 
yields
\begin{equation}
T_{00}^{\rm event}(\mathbf{k},\omega)
\simeq
-iE\,e^{-i\mathbf{k}\cdot\mathbf{X}}\,(v\,\mathbf{k}\cdot\hat{\mathbf{n}})\;
\tilde \chi(\omega),
\qquad
\tilde\chi(\omega)\equiv \int_{0}^{\tau_{\rm dip}} dt\, t\,e^{i\omega t}.
\label{eq:T00_event_k_omega}
\end{equation}
The time integral is elementary,
\begin{equation}
\tilde\chi(\omega)=
\frac{e^{i\omega\tau_{\rm dip}}(i\omega\tau_{\rm dip}-1)+1}{(i\omega)^2}.
\label{eq:chi_exact}
\end{equation}
Its squared magnitude has two useful limits:
\begin{equation}
|\tilde\chi(\omega)|^2 \simeq
\begin{cases}
\tau_{\rm dip}^4/4, & |\omega|\tau_{\rm dip}\ll 1,\\[4pt]
{\cal O}(1/\omega^4), & |\omega|\tau_{\rm dip}\gg 1,
\end{cases}
\label{eq:chi_limits}
\end{equation}
so the temporal spectrum is approximately flat at low frequency and rapidly suppressed above $\omega_c\sim 1/\tau_{\rm dip}$.
(For a top-hat ``instant dipole'' one instead obtains $|\tilde\chi|^2\propto \sin^2(\omega\tau_{\rm dip}/2)/\omega^2$; the growing dipole has an
additional $t$ weight that enhances low-frequency power by $\tau_{\rm dip}^2$.)

For a Poisson distribution of independent events with rate $\Gamma_{\rm dip}$ in spacetime, the connected two-point function is shot-noise-like:
\be
\left\langle T_{00}(\mathbf{k},\omega)\,T_{00}(\mathbf{k}',\omega')\right\rangle
=
(2\pi)^4\,\delta^{(3)}(\mathbf{k}+\mathbf{k}')\,\delta(\omega+\omega') P_{T_{00}}(\mathbf{k},\omega)
\ee
with
\be
P_{T_{00}}(\mathbf{k},\omega)\simeq \Gamma_{\rm dip}\,
\Big\langle |T_{00}^{\rm event}(\mathbf{k},\omega)|^2\Big\rangle_{\hat{\mathbf{n}}},
\label{eq:PT00_event_def}
\ee
where $\langle\cdots\rangle_{\hat{\mathbf{n}}}$ denotes the isotropic average over event orientations.
Using \eqref{eq:T00_event_k_omega}, we get
\begin{equation}
|T_{00}^{\rm event}(\mathbf{k},\omega)|^2
\simeq
E^2 v^2 (\mathbf{k}\cdot\hat{\mathbf{n}})^2\,|\tilde\chi(\omega)|^2.
\end{equation}
which for $|\omega|\tau_{\rm dip}\ll 1$ implies
\begin{equation}
P_{\Phi}(\mathbf{k},\omega)
\simeq
\frac{(4\pi G)^2}{3}\,
\Gamma_{\rm dip}\,E^2 v^2\;
\frac{|\tilde\chi(\omega)|^2}{k^2}.
\label{eq:Pphi_general}
\end{equation}
Therefore
on the low-frequency plateau,
\begin{equation}
P_{\Phi}(\mathbf{k},\omega)\ \simeq\
\frac{(4\pi G)^2}{12}\,
\Gamma_{\rm dip}\,\,E^2 v^2\;
\frac{\tau_{\rm dip}^4}{k^2}
\qquad (|\omega|\tau_{\rm dip}\ll 1).
\label{eq:Pphi_plateau_final}
\end{equation}
Note that unlike appendix B, $P_{\Phi}(k,\omega)$ here denotes the 4D spectral density, defined with a $(2\pi)^4\delta^{(3)}(k+k')\delta(\omega+\omega')$ factor; the corresponding equal-time 3D spectrum is obtained from
\[
P_{\Phi}^{\rm eq}(k)=\int \frac{d\omega}{2\pi}\,P_{\Phi}(k,\omega).
\]


\begin{thebibliography}{99}

\bibitem{Zeh1970}
H.~D.~Zeh,
``On the Interpretation of Measurement in Quantum Theory,''
Found.\ Phys.\ {\bf 1} (1970) 69--76.
doi:10.1007/BF00708656.

\bibitem{Mott1929}
N.~F.~Mott,
``The wave mechanics of $\alpha$-ray tracks,''
Proc.\ Roy.\ Soc.\ Lond.\ A {\bf 126} (1929) 79--84.
doi:10.1098/rspa.1929.0205.

\bibitem{Zurek2003}
W.~H.~Zurek,
``Decoherence, einselection, and the quantum origins of the classical,''
Rev.\ Mod.\ Phys.\ {\bf 75} (2003) 715--775.
doi:10.1103/RevModPhys.75.715.

\bibitem{Schlosshauer2005}
M.~Schlosshauer,
``Decoherence, the measurement problem, and interpretations of quantum mechanics,''
Rev.\ Mod.\ Phys.\ {\bf 76} (2005) 1267--1305.
doi:10.1103/RevModPhys.76.1267.

\bibitem{Everett1957}
H.~Everett, III,
``Relative State' Formulation of Quantum Mechanics,''
Rev.\ Mod.\ Phys.\ {\bf 29} (1957) 454--462.
doi:10.1103/RevModPhys.29.454.

\bibitem{GRW1986}
G.~C.~Ghirardi, A.~Rimini and T.~Weber,
``Unified dynamics for microscopic and macroscopic systems,''
Phys.\ Rev.\ D {\bf 34} (1986) 470--491.
doi:10.1103/PhysRevD.34.470.

\bibitem{Pearle1989}
P.~Pearle,
``Combining stochastic dynamical state-vector reduction with spontaneous localization,''
Phys.\ Rev.\ A {\bf 39} (1989) 2277--2289.
doi:10.1103/PhysRevA.39.2277.

\bibitem{GPR1990}
G.~C.~Ghirardi, P.~Pearle and A.~Rimini,
``Markov processes in Hilbert space and continuous spontaneous localization of systems of identical particles,''
Phys.\ Rev.\ A {\bf 42} (1990) 78--89.
doi:10.1103/PhysRevA.42.78.

\bibitem{Diosi1989}
L.~Di\'osi,
``Models for universal reduction of macroscopic quantum fluctuations,''
Phys.\ Rev.\ A {\bf 40} (1989) 1165--1174.
doi:10.1103/PhysRevA.40.1165.

\bibitem{Penrose1996}
R.~Penrose,
``On gravity's role in quantum state reduction,''
Gen.\ Rel.\ Grav.\ {\bf 28} (1996) 581--600.
doi:10.1007/BF02105068.

\bibitem{BassiGhirardi2003}
A.~Bassi and G.~C.~Ghirardi,
``Dynamical reduction models,''
Phys.\ Rept.\ {\bf 379} (2003) 257--426.
doi:10.1016/S0370-1573(03)00103-0.

\bibitem{BassiEtAl2013}
A.~Bassi, K.~Lochan, S.~Satin, T.~P.~Singh and H.~Ulbricht,
``Models of wave-function collapse, underlying theories, and experimental tests,''
Rev.\ Mod.\ Phys.\ {\bf 85} (2013) 471--527.
doi:10.1103/RevModPhys.85.471.

\bibitem{BassiDoratoUlbricht2023}
A.~Bassi, M.~Dorato and H.~Ulbricht,
``Collapse Models: A Theoretical, Experimental and Philosophical Review,''
Entropy \textbf{25}, no.4, 645 (2023)
doi:10.3390/e25040645
[arXiv:2310.14969 [quant-ph]].

%\cite{Fein:2019dgf}
\bibitem{Fein:2019dgf}
Y.~Y.~Fein, P.~Geyer, P.~Zwick, F.~Kia{\l}ka, S.~Pedalino, M.~Mayor, S.~Gerlich and M.~Arndt,
``Quantum superposition of molecules beyond 25 kDa,''
Nature Phys. \textbf{15}, no.12, 1242-1245 (2019)
doi:10.1038/s41567-019-0663-9
%244 citations counted in INSPIRE as of 19 Mar 2026

%\bibitem{GerlichFeinArndt2021}
%S.~Gerlich, Y.~Fein and M.~Arndt,
%``Interferometric tests of wave-function collapse,''
%in V.~Allori, A.~Bassi, D.~D\"urr and N.~Zangh\`i eds.,
%{\it Do Wave Functions Jump?: Perspectives of the Work of GianCarlo Ghirardi},
%Fundamental Theories of Physics %{\bf 198} (2021) 385--399.
%doi:10.1007/978-3-030-46777-7_26.

\bibitem{CarlessoPaternostro2020}
M.~Carlesso and M.~Paternostro,
``Opto-mechanical tests of collapse models,''
Fundam. Theor. Phys. \textbf{198}, 205-215 (2020)
doi:10.1007/978-3-030-46777-7{\_}16
[arXiv:1906.11041 [quant-ph]].
%6 citations counted in INSPIRE as of 19 Mar 2026M.~Carlesso and M.~Paternostro,


\bibitem{VinanteEtAl2016}
%\cite{Vinante:2015xxe}
%\bibitem{Vinante:2015xxe}
A.~Vinante, M.~Bahrami, A.~Bassi, O.~Usenko, G.~Wijts and T.~H.~Oosterkamp,
``Upper Bounds on Spontaneous Wave-Function Collapse Models Using Millikelvin-Cooled Nanocantilevers,''
Phys. Rev. Lett. \textbf{116}, no.9, 090402 (2016)
doi:10.1103/PhysRevLett.116.090402
[arXiv:1510.05791 [quant-ph]].


\bibitem{CarlessoEtAl2022}
%\cite{Carlesso:2022pqr}
%\bibitem{Carlesso:2022pqr}
M.~Carlesso, S.~Donadi, L.~Ferialdi, M.~Paternostro, H.~Ulbricht and A.~Bassi,
``Present status and future challenges of non-interferometric tests of collapse models,''
Nature Phys. \textbf{18}, no.3, 243-250 (2022)
doi:10.1038/s41567-021-01489-5
[arXiv:2203.04231 [quant-ph]].


\bibitem{Fu1997}
Q.~Fu,
``Spontaneous radiation of free electrons in a nonrelativistic collapse model,''
Phys.\ Rev.\ A {\bf 56} (1997) 1806--1811.
doi:10.1103/PhysRevA.56.1806.

\bibitem{ArnquistEtAl2022}
%\cite{Majorana:2022xuq}
%\bibitem{Majorana:2022xuq}
I.~J.~Arnquist \textit{et al.} [Majorana],
``Search for Spontaneous Radiation from Wave Function Collapse in the Majorana Demonstrator,''
Phys. Rev. Lett. \textbf{129}, no.8, 080401 (2022)
[erratum: Phys. Rev. Lett. \textbf{130}, no.23, 239902 (2023)]
doi:10.1103/PhysRevLett.129.080401
[arXiv:2202.01343 [nucl-ex]].


%\cite{XENON:2025plg}
\bibitem{XENON:2025plg}
E.~Aprile \textit{et al.} [XENON],
``Challenging Spontaneous Quantum Collapse with XENONnT,''
[arXiv:2506.05507 [hep-ex]].
%1 citations counted in INSPIRE as of 21 Mar 2026


\bibitem{Maldacena1998}
J.~M.~Maldacena,
``The Large $N$ Limit of Superconformal Field Theories and Supergravity,''
Adv.\ Theor.\ Math.\ Phys.\ {\bf 2} (1998) 231--252
[Int.\ J.\ Theor.\ Phys.\ {\bf 38} (1999) 1113--1133].
doi:10.1023/A:1026654312961.

\bibitem{GKP1998}
S.~S.~Gubser, I.~R.~Klebanov and A.~M.~Polyakov,
``Gauge Theory Correlators from Non-Critical String Theory,''
Phys.\ Lett.\ B {\bf 428} (1998) 105--114.
doi:10.1016/S0370-2693(98)00377-3.

\bibitem{Witten1998}
E.~Witten,
``Anti De Sitter Space and Holography,''
Adv.\ Theor.\ Math.\ Phys.\ {\bf 2} (1998) 253--291.
doi:10.4310/ATMP.1998.v2.n2.a2.

\bibitem{Witten2001}
E.~Witten,
``Quantum Gravity in de~Sitter Space,''
hep-th/0106109.

\bibitem{DanielssonVanRiet2018}
U.~H.~Danielsson and T.~Van Riet,
``What if String Theory has no de~Sitter Vacua?,''
Int.\ J.\ Mod.\ Phys.\ D {\bf 27} (2018) 1830007.
doi:10.1142/S0218271818300070.

\bibitem{DineEtAl2021}
M.~Dine, J.~A.~P.~Law-Smith, S.~Sun, D.~Wood and Y.~Yu,
``Obstacles to constructing de~Sitter space in string theory,''
JHEP {\bf 02} (2021) 050.
doi:10.1007/JHEP02(2021)050.

\bibitem{CicoliEtAl2024}
M.~Cicoli, J.~P.~Conlon, A.~Maharana, S.~Parameswaran, F.~Quevedo and I.~Zavala,
``String Cosmology: from the Early Universe to Today,''
Phys.\ Rept.\ {\bf 1059} (2024) 1--155.
doi:10.1016/j.physrep.2024.01.002.

\bibitem{Itzhaki:2024pok}
N.~Itzhaki and U.~Peleg,
``Instant cosmology,''
JHEP {\bf 05} (2025) 026.
doi:10.1007/JHEP05(2025)026
[arXiv:2412.02630 [hep-th]].

\bibitem{Itzhaki:2018glf}
N.~Itzhaki,
``Stringy instability inside the black hole,''
JHEP {\bf 10} (2018) 145.
doi:10.1007/JHEP10(2018)145
[arXiv:1808.02259 [hep-th]].

\bibitem{Penrose1998}
R.~Penrose,
``Quantum computation, entanglement and state reduction,''
Phil.\ Trans.\ Roy.\ Soc.\ Lond.\ A {\bf 356} (1998) 1927--1939.
doi:10.1098/rsta.1998.0256.

\bibitem{Diosi1987}
L.~Di\'osi,
``A universal master equation for the gravitational violation of quantum mechanics,''
Phys.\ Lett.\ A {\bf 120} (1987) 377--381.

\bibitem{Donadi2021}
S.~Donadi, K.~Piscicchia, C.~Curceanu, L.~Di\'osi, M.~Laubenstein and A.~Bassi,
``Underground test of gravity-related wave function collapse,''
Nature Phys.\ {\bf 17} (2021) 74--78.
doi:10.1038/s41567-020-1008-4.

\bibitem{Diosi2014Bulk}
L.~Di\'osi,
``Gravity-related spontaneous collapse in bulk matter,''
arXiv:1404.6644 [quant-ph].

%\bibitem{BassiRMP2013}
%A.~Bassi, K.~Lochan, S.~Satin, T.~P.~Singh and H.~Ulbricht,
%``Models of wave-function collapse, underlying theories, and experimental tests,''
%Rev.\ Mod.\ Phys.\ {\bf 85} (2013) 471--527.
%doi:10.1103/RevModPhys.85.471.

%\bibitem{CarlessoReview}
%M.~Carlesso et al.,
%``Present status and future challenges of non-interferometric tests of collapse models,''
%arXiv:2203.04231 [quant-ph].

%\bibitem{Figurato2024}
%\cite{Figurato:2024tpo}
\bibitem{Figurato:2024tpo}
L.~Figurato, M.~Dirindin, J.~L.~Gaona-Reyes, M.~Carlesso, A.~Bassi and S.~Donadi,
``On the effectiveness of the collapse in the Di{\'o}si{\textendash}Penrose model,''
New J. Phys. \textbf{26}, no.11, 113004 (2024)
doi:10.1088/1367-2630/ad8c77
[arXiv:2406.18494 [quant-ph]].
%10 citations counted in INSPIRE as of 15 Mar 2026
%A.~Figurato, A.~Bassi and collaborators,
%``On the effectiveness of the collapse in the Di\'osi--Penrose model,''
%arXiv:2406.18494 [quant-ph].

\bibitem{AdlerBassi2007}
S.~L.~Adler and A.~Bassi,
``Collapse models with non-white noises,''
J.\ Phys.\ A {\bf 40} (2007) 15083--15098.
doi:10.1088/1751-8113/40/50/012.

\bibitem{CarlessoFerialdiBassi2018}
M.~Carlesso, L.~Ferialdi and A.~Bassi,
``Colored collapse models from the non-interferometric perspective,''
Eur.\ Phys.\ J.\ D {\bf 72} (2018) 159.
doi:10.1140/epjd/e2018-90248-x.

\bibitem{Helou2017LPF}
B.~Helou, B.~Slagmolen, D.~E.~McClelland and Y.~Chen,
``LISA Pathfinder appreciably constrains collapse models,''
Phys.\ Rev.\ D {\bf 95} (2017) 084054.
doi:10.1103/PhysRevD.95.084054.

\bibitem{Riess1998}
A.~G.~Riess et al.,
``Observational Evidence from Supernovae for an Accelerating Universe and a Cosmological Constant,''
Astron.\ J.\ {\bf 116} (1998) 1009--1038.
doi:10.1086/300499.

\bibitem{Perlmutter1999}
S.~Perlmutter et al.,
``Measurements of Omega and Lambda from 42 High-Redshift Supernovae,''
Astrophys.\ J.\ {\bf 517} (1999) 565--586.
doi:10.1086/307221.

\bibitem{Attali:2018goq}
K.~Attali and N.~Itzhaki,
``The Averaged Null Energy Condition and the Black Hole Interior in String Theory,''
Nucl.\ Phys.\ B {\bf 943} (2019) 114631.
doi:10.1016/j.nuclphysb.2019.114631
[arXiv:1811.12117 [hep-th]].

\bibitem{ItzhakiPeleg2024}
N.~Itzhaki and U.~Peleg,
``When Strings Surprise,''
JHEP {\bf 08} (2024) 172.
doi:10.1007/JHEP08(2024)172
[arXiv:2404.03215 [hep-th]].

%\cite{Hashimoto:2022dro}
\bibitem{Hashimoto:2022dro}
A.~Hashimoto, N.~Itzhaki and U.~Peleg,
``A worldsheet description of instant folded strings,''
JHEP \textbf{02}, 088 (2023)
doi:10.1007/JHEP02(2023)088
[arXiv:2209.04988 [hep-th]].
%9 citations counted in INSPIRE as of 22 Mar 2026
%\cite{Itzhaki:2023hwm}
\bibitem{Itzhaki:2023hwm}
N.~Itzhaki,
``Is the horizon of an eternal black hole really smooth?,''
JHEP \textbf{05}, 157 (2023)
doi:10.1007/JHEP05(2023)157
[arXiv:2302.03556 [hep-th]].
%1 citations counted in INSPIRE as of 24 Mar 2026
%\cite{Strominger:1990et}
\bibitem{Strominger:1990et}
A.~Strominger,
``Heterotic solitons,''
Nucl. Phys. B \textbf{343}, 167-184 (1990)
[erratum: Nucl. Phys. B \textbf{353}, 565-565 (1991)]
doi:10.1016/0550-3213(90)90599-9
%362 citations counted in INSPIRE as of 15 Mar 2026

%\cite{Maldacena:1998im}
\bibitem{Maldacena:1998im}
J.~M.~Maldacena,
``Wilson loops in large N field theories,''
Phys. Rev. Lett. \textbf{80}, 4859-4862 (1998)
doi:10.1103/PhysRevLett.80.4859
[arXiv:hep-th/9803002 [hep-th]].
%2143 citations counted in INSPIRE as of 18 Mar 2026

%\cite{Rey:1998ik}
\bibitem{Rey:1998ik}
S.~J.~Rey and J.~T.~Yee,
``Macroscopic strings as heavy quarks in large N gauge theory and anti-de Sitter supergravity,''
Eur. Phys. J. C \textbf{22}, 379-394 (2001)
doi:10.1007/s100520100799
[arXiv:hep-th/9803001 [hep-th]].
%1534 citations counted in INSPIRE as of 18 Mar 2026


%\cite{tHooft:1993dmi}
\bibitem{tHooft:1993dmi}
G.~'t Hooft,
``Dimensional reduction in quantum gravity,''
Conf. Proc. C \textbf{930308}, 284-296 (1993)
[arXiv:gr-qc/9310026 [gr-qc]].
%3473 citations counted in INSPIRE as of 18 Mar 2026

%\cite{Susskind:1994vu}
\bibitem{Susskind:1994vu}
L.~Susskind,
``The World as a hologram,''
J. Math. Phys. \textbf{36}, 6377-6396 (1995)
doi:10.1063/1.531249
[arXiv:hep-th/9409089 [hep-th]].
%4361 citations counted in INSPIRE as of 18 Mar 2026

\bibitem{PasterskiShaoStrominger2017}
S.~Pasterski, S.-H.~Shao and A.~Strominger,
``Flat Space Amplitudes and Conformal Symmetry of the Celestial Sphere,''
Phys.\ Rev.\ D {\bf 96} (2017) 065026
[arXiv:1701.00049 [hep-th]].

\bibitem{MeltonSharmaStromingerWang2024}
W.~Melton, A.~Sharma, A.~Strominger and T.~Wang,
``A Celestial Dual for MHV Amplitudes,''
[arXiv:2403.18896 [hep-th]].


\bibitem{Itzhaki:2025gdv}
N.~Itzhaki, U.~Peleg and P.~J.~Steinhardt,
``Instant Folded Strings, Dark Energy and a Cyclic Bouncing Universe,''
arXiv:2508.09745 [gr-qc].




\bibitem{SteinhardtTurok2002PRD}
P.~J.~Steinhardt and N.~Turok,
``Cosmic Evolution in a Cyclic Universe,''
{\it Phys.\ Rev.\ D} {\bf 65} (2002) 126003
[arXiv:hep-th/0111098].



\bibitem{IjjasSteinhardt2019NewCyclic}
A.~Ijjas and P.~J.~Steinhardt,
``A New Kind of Cyclic Universe,''
{\it Phys.\ Lett.\ B} {\bf 795} (2019) 666--672
[arXiv:1904.08022 [gr-qc]].

\bibitem{IjjasSteinhardt2022Entropy}
A.~Ijjas and P.~J.~Steinhardt,
``Entropy, Black Holes, and the New Cyclic Universe,''
{\it Phys.\ Lett.\ B} {\bf 824} (2022) 136823
[arXiv:2108.07101 [gr-qc]].


\bibitem{PeeblesLSS1980}
P.~J.~E.~Peebles,
\emph{The Large-Scale Structure of the Universe},
Princeton University Press (1980).



\end{thebibliography}
\end{document}